# BAGH – COMPARATIVE STUDY

B. Kamala
Assistant Professor, Department of Information Technology
Sri Sai Ram Engineering College, Chennai – 44.
kamala.it@sairam.edu.in

*Abstract* – Process mining is a new emerging research trend over the last decade which focuses on analyzing the processes using event log and data. The raising integration of information systems for the operation of business processes provides the basis for innovative data analysis approaches. Process mining has the strong relationship between with data mining so that it enables the bond between business intelligence approach and business process management. It focuses on end-to-end processes and is possible because of the growing availability of event data and new process discovery and conformance checking techniques. Process mining aims to discover, monitor and improve real processes by extracting knowledge from event logs readily available in today's information systems. The discovered process models can be used for a variety of analysis purposes. Many companies have adopted Process-aware Information Systems for supporting their business processes in some form. These systems typically have their log events related to the actual business process executions. Proper analysis of Process Aware Information Systems execution logs can yield important knowledge and help organizations improve the quality of their services. This paper reviews and compares various process mining algorithms based on their input parameters, the techniques used and the output generated by them.

## I. INTRODUCTION

Process mining is applicable to a wide range of systems. These systems may be pure information systems or systems where the hardware plays a more vital role. The only requirement of that system is that it has to produce an event log which records the actual behavior of the system. Process-Aware Information Systems are the systems that produce interesting classes of information systems that produce event logs. By analyzing these systems execution logs can bring knowledge about the organizations and help organizations to improve the quality of their services. These systems provide very detailed information about the activities that have been executed.[1][15] The goal of process mining is to extract information (e.g., process models) from these logs, i.e., process mining describes a family of a-posteriori analysis techniques exploiting the information recorded in the event logs. Process mining addresses the problem that most process or system owners have limited information about what is actually happening. Process mining builds the connection between data mining as a business intelligence approach and business process management. [6]

The remainder of this paper is organized as follows: Section 2 describes the scope of data mining, Section 3 depicts the scope of process mining, Section 4 compares Data Mining and Process Mining and Section 5 encompasses of the comparative study of various process mining algorithms.

## II. What is Data Mining?

Discovering patterns and finding facts and truth from large data sets is the process of data mining. Extracting information from a large data set and the extracted information is converted into an understandable structure is the goal of a data mining process. DM process involves inventing and finding valid, new, useful and logically understandable data. These data predicts a regularity among the data variables available in the source. If the regularity applies to all data variables, then it is a discovered model and if it is correlated with some extent of data, then it a template or pattern.

The model discovery or a pattern identification can be carried out over the large volume of data, so that a better business decision can be made. These discovery models are used to find knowledge and these knowledge can be represented through patterns[2]. The most commonly used method in data mining is association rule which is used to find the relationship between data and various objects. This association rule depicts dependencies among various data and objects. Clustering and classification can also be used to find common significances among various objects [9].

Data mining in a centralized databases have the drawbacks: less network traffic, low mining efficiency and spatial complexity is relatively high. Classification based on distance, Decision tree induction, Naïve Bayesian classifiers are the traditional classification methods [7]. Various clustering can also be used to generate a pattern or template to present a knowledge. Visualization tools can also be used to visualize the data patterns in a graphical format.

### A. Data Mining parameters
The data mining parameters are as follows:
- Association – provides the relationship between data variables and other related objects.
- Path analysis – depicts whether one event leads to another event
- Classification – Generating new patterns.
- Predictive analysis – predictions about the future by discovering patterns
- Clustering – unsupervised learning method where facts are unknown

*B. Steps involved in Data Mining process*

The steps involved in the data mining process are,
- Definition of the business problem – Business problem is defined and collect the required data
- Data preparation – Prepare the data by data transformation, data sampling and data evaluation
- Modeling – Select a proper mining technique, build a model and evaluate it.
- Implementation – Implement the model and deploy it

The following figure explains the different steps which comprise the overall data mining process.

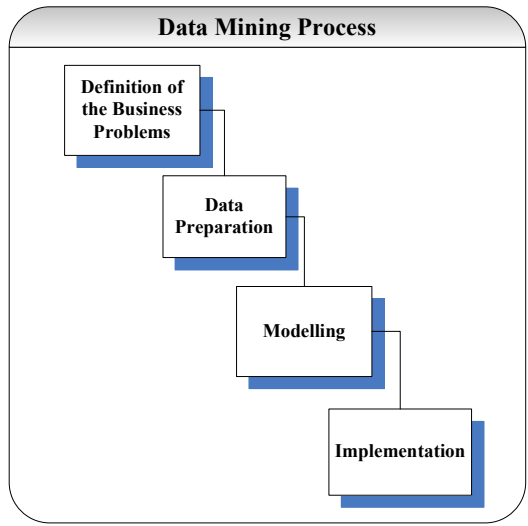

Fig.1 Data Mining Process

### III. SCOPE OF PROCESS MINING MODELS

Process models are constructed based on the logging data using process mining. Process models are used to visualize the business process describing the relationship between business processes for a particular business perspective. There are three categories in process mining model. They are Discovery, Conformance and Enhancement.

Discovery – Information about the original process model, context of the organization and execution properties from process logs are focused in process discovery.

Conformance – Detects deviations, locate and explain these deviations. It is an apiori model and used to match the reality with model. This is also used to measure the severity of the deviations.

Enhancement – It is also an apriori model. This model extends the new aspect or the perspective about the model. This model enriches the existing model with the data in the event log [4].

The following figure shows the architecture of Process Mining. The real world gives the raw events and runtime data to the process state and history repository.

[13] From that repository Process mining algorithms traces for each process instances so that it can generate models and analysis from the process state repository.

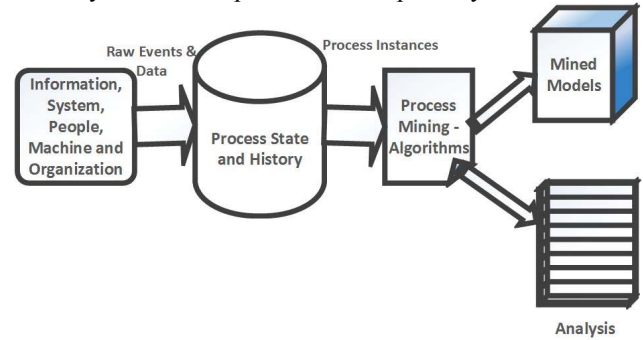

Fig.2 Process Mining Architecture

The existing model is compared with the new derived model. Conformance checking is used to check the recorded logs are implemented in a right manner and enhancement is used to give suggestion or betterment to improve the existing model. [10]

*A. Discovery Model*

Discovery model produces a new model by taking an event log as its input without any prior information. Process discovery is one of the most important step in process mining technique. In general, the behavior of an organization is stored as event log. The organization can able to discover the real process by discovering a process model from their event logs. This model automatically constructs model based on the event logs and observed events. [3]

The discovered process models may be used
- Problem discussion with stakeholders
- Generate process improvement ideas
- Model enhancement and analyze bottleneck
- Can be served as a template

*B. Conformance Model*

Conformance checking model constructs a model based on the event logs generated while discovery constructs a model with prior information. This model is used to relate events in the logs with activities. This model is used compare the modelled behavior with the observed behavior..

Conformance checking can be used for
- Compare modelled behavior with observed behavior
- Deviating cases are identified
- Discovered process model quality is judged
- Monitoring the process of discovery algorithms
- Initiation for model enhancement

*C. Enhancement Model*

An event log is used to improve or extend the existing process model. By analyzing the association between a model and a log, a process model can be generated or modified. Event logs will contain the information about the resources used by the process, the timestamps, and relevant data. Information about these resources are used to identify the roles of processes. Constructing social networks is also possible by analyzing the work flow and resource performance.

*D. Process Models and Event Log*

Minimal information to represent a process in log uses their case id and task id. Additional information may require will be its event type, time, resources, and data. The following figure and table describes a case study of event log which contains case id and event id timestamp and its associated activity.

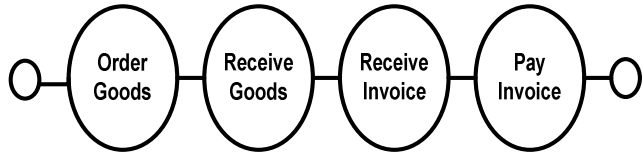

| CASE ID | EVENT ID | TIMESTAMP | ACTIVITY |
|---|---|---|---|
| 1 | 1000 | 01.10.2018 | Order Goods |
|  | 1001 | 10.10.2018 | Receive Goods |
|  | 1002 | 13.10.2018 | Receive Invoice |
|  | 1003 | 20.10.2018 | Pay Invoice |
| 2 | 1004 | 02.11.2018 | Order Goods |
|  | 1005 | 11.11.2018 | Receive Goods |
|  | …… | …… | …… |

Fig. 3 Process Model and an Event Log

## IV. DATA MINING VS PROCESS MINING

Apart from mining part, there are many similarities between process mining and data mining. Process mining takes the large amount and volume of data like data mining. Business Intelligence is a common term used in both data mining and process mining. Business Intelligence retrieves valuable knowledge from the analysis of a large volume of digital data which is available from mining tools and techniques. Business Intelligence can be either derived by data mining or of by process mining. But the derived business intelligence is used to take many business solutions.

Process mining bridges data mining, big data and business process management which is described in the following figure.

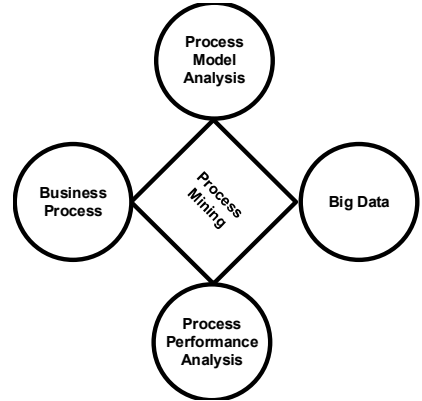

Fig. 4 Role of Process Mining

Process mining technology associates the strength of both process modelling and data mining. Process mining brings live models by automatically creating process models based on the existing IT log data. These models are always connected with business and can be updated at any point of time. Since process models are generated manually most of the times, they will be outdated. Model enhancement will by little bit difficult by using the outdated process models. So, the following guidelines can be used to generate a process model.

- Treat Event data as an important identity
- Extraction of logs will be always based on questionnaires
- Follow concurrency mechanisms, choice control and other basic control flow constructs
- Treat events as model elements
- Treat models as reality abstractions
- Perform Process mining as a continuous process

The following table describes the similarities and differences between process mining and data mining in various perspectives like data volume, outcome, goal and the model generated by the algorithm.

TABLE 1 DATA MINING VS PROCESS MINING

| Perspective | Data Mining | Process Mining |
|---|---|---|
| Objective of the Model | Data mining techniques are pattern perspective and it is used to find the relationship between the data patterns | Process mining focusses mainly on processes. It also includes the temporal aspect of a process. Process execution can be treated as sequence of activities |
| Volume of data | Use large volume of data | Use large volume of data |
| Outcome | Used for making business decisions | Used for making business decisions |
| Goal | Fact finding and association between data | Model discovery and finds associations between data and |

|  | patterns | event logs |
|---|---|---|
| Generated Model | Generated model is always in the form of rules or decision trees | Generated model is always a process model which describes associations between data and event logs |

## V. ALGORITHMS USED IN PROCESS MINING

Process mining algorithms are the major components of process mining. It is used to determine how the models are created. The following are the major categories of process mining algorithms.

- Bit-coin mining algorithm
- Alpha algorithm
- Genetic mining algorithm
- Heuristic mining algorithm

### A. Bit-coin mining algorithm

Transactional databases are used to mine the bit-coins that are generated in the bit-coin economy. Bit-coin mining algorithm maintains a single transaction history in which it prevents bit-coins being spent doubly. This algorithm uses a technique called block, which collects all transactions during a specific period. Miner has to confirm these transactions and they have to keep a record. These records are used to identify all transactions across network. If there is a new transaction occurred in a network, then it will be added in the blockchain. A blockchain consists of list of transactions that are occurred in a bit-coin network. The miners have to update the variance called nonce, which is a 32bit value which updates the hash value of the contents of block in blockchain. Updated copy of this block will be given to all miners in the bit-coin network, so everyone will be aware of it.[14] Steps involved in bit-coin mining algorithm is as follows:

- Block header will be given as input
- Modify Nonce
- If block header hash is less than target, then you win
- Else, Repeat steps 1 and 2 since someone won the block

### B. Alpha algorithm

Alpha algorithm is a process mining algorithm used to reconstruct the association between the set of sequences of events. One of the possible solution of alpha algorithm is a process model which extracts workflow models from event logs. A workflow in alpha algorithm is a collection of workflow traces. A workflow trace or execution trace in this algorithm generates a string over an alphabet. [14] The steps involved in Alpha algorithm is as follows:

- Represent the transition as a collection of atomic activities
- Select single state and end phase
- Infer workflow trace and workflow log
- Find log-based ordering relations using basic relations

### C. Genetic mining algorithm

Genetic mining algorithm is a traditional way of approach to generate process models. This algorithm duplicates the process of natural evolution. A random population of process models has been generated and from these models finds a solution by iteratively selecting individual model by crossover and mutation process. The initial population generated by this algorithm is slightly similar to that of the event log. Genetic algorithm uses representations, mutations, crossovers and selection mechanisms in its process. [14]Genetic mining algorithm includes the following steps.

- Generate initial population
- Apply initial population for fitness
- Select best parents
- Apply cross over and mutation methods for generation of new children
- New population is generated

### D. Heuristic mining algorithm

Heuristic mining algorithm is an applied mining algorithm which deals with noise and is used to express the major behavior that is represented by the event log. This algorithm mines the control flow perspective of a process model. Order of events in the initial process is considered by this algorithm. Parameters used for mining process logs are case id, timestamp and all associated activities. By analyzing casual dependencies of event logs, process model can be derived. If one activity follows another activity, then it is found that there is a dependency relation between those two activities[14] [16]. The heuristic mining algorithm involves the following steps:

- Dependency graph to be constructed
- Indicate frequency based metric which is used to indicate dependency relation between two events
- Mine the dependency graph
- Define the dependency relations for non-observable tasks by constructing the casualty matrix
- Mine long distance dependencies

The following table describes the type of process mining algorithm, techniques used in that algorithm and the output generated by the algorithm. [8]

**TABLE II COMPARATIVE STUDY ON PROCESS MINING ALGORITHMS**

| Process Mining Algorithms | Bit-coin mining Algorithm | Alpha Algorithm | Genetic Algorithm | HeuristicMiner Algorithm |
|---|---|---|---|---|
| Algorithm Type | Deterministic mining algorithm | Deterministic mining algorithm | Genetic algorithm | Heuristic algorithm |
| Basic concept | A block is used, which collects all transactions during a specific period. Miner has to confirm these transactions and they have to keep a record. | Alpha algorithm is a process mining algorithm used to reconstruct the association between the set of sequences of events. | Genetic algorithms are categorized as global search heuristics | Used to represent the major behavior that is represented by the event log |
| Techniques used | Block, General ledger and Block chain | Sequence of events and Event Log | Representations, mutations, crossovers and Selection mechanisms | Behavior registered in event log, dependency relationship between activities |
| Strategy | Maintains a single transaction history in which it prevents bit-coins being spent doubly | Local strategy technique to build model | A single solution is carried over from one iteration to the next iteration | Work flow model, frequency based metric |
| Use when | When there is a need of blocks and block chains to generate process models | When there is a set of workflow sequences | When there is a need of natural evolution of process model using cross over and mutation | When there is a need of representing major behavior of the process model by eliminating noise. |
| Issues | Nonce updation and human record maintenance | Information gathering from variety of sources, data visualization and analysis | Technique can deal with noisy and duplicate tasks. | Finding out the casual dependencies between any two activities is little bit difficult |
| Behavior | Infer patterns from blockchain | Infer workflow model from event samples | Geneticmining algorithmsmimicnaturalevolution | Heuristicminingalgorithmstakefrequenciesinto account |
| Output | Definedand ReproducibleResults | Work Flow Model | MimicNatural Evolution | Frequencies of the events |

## VI. CONCLUSION

This paper presents a comparative study on various process mining algorithms with respect to its techniques and the output generated by the algorithms. Process mining has the strong relationship between with data mining so that it enables the bond between business intelligence approach and business process management. Process mining mainly focuses on end to end processes which takes the help of all the three process mining models viz, process discovery, conformance and enhancement. Analysis of these process aware information systems using event logs helps to improve organizations quality of service. Thus, from this comparison it is very clear that comparing process mining algorithms with a collection of business models will be very expensive, tough and time consuming. Comparison between various process mining algorithms is carried out here based on their input parameters, the techniques used and the output generated by them.

Modelling business events and logs using process mining algorithm provides new type in data analysis. While mining process models, reconstruction of association rules, correlation between facts can also be identified. In future, a framework for comparing various process mining algorithms can be provided for process discovery against a known ground truth, with an implementation using existing tools.